 \documentclass[reprint, amsmath,amssymb,aps, prx,superscriptaddress]{revtex4-2}
 
\usepackage{ulem}
\usepackage{graphicx}
\usepackage{dcolumn}
\usepackage{bm}
\usepackage{xcolor}
\usepackage[dvipsnames]{xcolor}
\usepackage{amsmath}
\usepackage{amssymb}
\usepackage{lineno}
\usepackage{physics, mathtools}


\usepackage{hyperref}
\hypersetup{colorlinks=true, citecolor=blue, linkcolor=black, urlcolor=blue}

\newcommand{\pratik}[1]{\textcolor{magenta}{#1}}

\newcommand{\aurelia}[1]{\textcolor{Aquamarine}{#1}}
\newcommand{\niklas}[1]{\textcolor{ForestGreen}{#1}}

\begin{document}

\title{Measurement and Control of the Complex Berry Phase in a Quantum System}

\author{Pratik J. Barge}
\altaffiliation[These authors contributed equally to this work.]{}
\affiliation{Department of Physics, Washington University, St. Louis, Missouri 63130, USA}
\author{Qian Cao}
\altaffiliation[These authors contributed equally to this work.]{}
\affiliation{Department of Physics, Washington University, St. Louis, Missouri 63130, USA}
\author{Niklas H\"ornedal}
\author{Aur\'elia Chenu}
\affiliation{Department of Physics and Materials Science, University of Luxembourg, L-1511 Luxembourg, G. D. Luxembourg}
\author{Kater W. Murch}%
    \email{katermurch@berkeley.edu}
    \affiliation{Department of Electrical Engineering and Computer Science, University of California Berkeley, Berkeley, CA, USA, 94720.}
    \affiliation{Department of Physics, University of California Berkeley, Berkeley, CA, USA, 94720}

\date{\today}

             
\begin{abstract}
The Berry phase is a geometric phase acquired during adiabatic evolution over a closed loop in parameter space. It plays an essential role in geometric quantum gates and other phase-based protocols. In non-Hermitian systems, the Berry phase is complex, introducing fundamentally new geometric effects, including state amplification. In this work, we report experimental measurement of both the real and imaginary components of a Berry phase in a fully quantum system using a superconducting transmon circuit with engineered dissipation. We also demonstrate the path-dependent effects of the imaginary part on the dissipation and its utility in the implementation of non-unitary quantum control. These findings establish a clear geometric distinction between the real and imaginary components of the Berry phase and experimentally confirm the unique adiabatic behavior of non-Hermitian quantum systems.  
\end{abstract} 

\maketitle

\section{Introduction}
The Berry phase is a foundational concept in adiabatic quantum evolution. In Hermitian systems, it is a real-valued phase factor acquired under a closed adiabatic trajectory. The Berry phase, and more generally, geometric phases, have broad implications across quantum science: They underlie universal geometric quantum gates with intrinsic robustness to control errors~\cite{zhu2002implementation, abdumalikov2013experimental, zhou2025high}. Geometric phases govern measurement-induced phase phenomena~\cite{cho2019emergence, wang2022observing}. They also support sensing and simulation protocols across condensed-matter platforms, ranging from graphene quantum Hall systems~\cite{zhang2005experimental} and phase-based magnetometry~\cite{arai2018geometric} to synthetic gauge fields in ultra-cold atoms~\cite{lin2009synthetic}.

Non-Hermitian (NH) Hamiltonians, originally developed to model open quantum dynamics, have become a powerful framework for describing effective evolution in a variety of classical as well as quantum experimental platforms \cite{bender2013observation, ruter2010observation, schindler2012symmetric, zhu2014pt, Xu2016, ozdemir2019parity, ding2021experimental, Nagh19}.
NH systems exhibit unique (complex-valued) spectral and eigenstate properties absent in closed quantum systems, such as exceptional point (EP) degeneracies, non-orthogonal eigenkets, and chiral state transfer. Adiabatic evolution in such systems acquires a fundamentally different character and leads to a complex-valued Berry phase. The real part adds to the Berry phase of closed quantum systems, while the imaginary part, which encodes the anti-Hermitian part of the Hamiltonian, arises from state-dependent amplification or attenuation. Recent experiments have observed the complex Berry phase using classical platforms, such as feedback-coupled oscillators ~\cite{singhal2023measuring} and an optomechanical system~\cite{lane2025complex}. 
Theoretical work has investigated the extension of the complex Berry phase to NH many-body systems, predicting quantization behavior~\cite{tsubota2022symmetry}. 

In this work, we measure both the real and imaginary components of the complex geometric phase arising from adiabatic evolution of the quantum states of a superconducting transmon circuit coupled to an engineered dissipative channel. We observe path-dependent amplification or attenuation of eigenstates, corresponding to the imaginary component of the geometric phase. The characterization of the imaginary geometric phase unlocks  tunable, non-unitary control for geometric manipulation of quantum states.

This paper is organized as follows: Section \ref{sec:model} introduces the theoretical framework for non-Hermitian adiabatic evolution and the definition of the complex Berry phase. Section \ref{sec:experiment} describes our experimental platform and the measurement protocols. The distinct physical roles of the real and imaginary components are analyzed within these sections. Section \ref{sec:nonunitary} shows how to use the Berry phase for non-unitary control. Section \ref{sec:conclude} provides concluding remarks. 


\section{Model \label{sec:model}}
\begin{figure}[t]
  \centering
  \includegraphics[width=1\columnwidth]{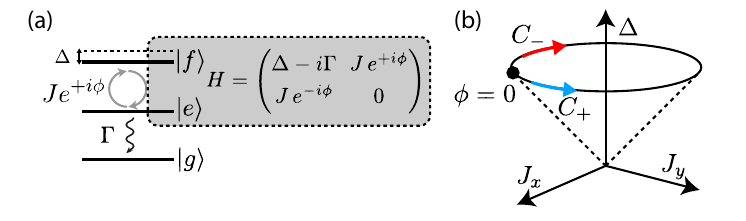}
  \caption{(a) An effective non-Hermitian system is realized through post-selection on trajectories with no quantum jumps from $\ket{e}$ to $\ket{g}$ at decay rate $\Gamma$, while the $\ket{e}$–$\ket{f}$ manifold is coherently driven with amplitude $J$ and detuning $\Delta$. To enable high-fidelity state readout, population in $\ket{e}$ is transferred to a neighboring qubit using a parametric SWAP operation.
  (b) Closed parameter-space loops $C_\pm$ traced in $\{J_x, J_y, \Delta\}$ during the geometric evolution 
  }  
  \label{fig:fig1}
\end{figure}

In this work, we focus on the dynamics of a NH qubit, shown in Fig.~\ref{fig:fig1}a. As detailed in the next Section, the Hamiltonian is generated by driving a qubit with detuning $\Delta$ and amplitude $J$ leading to an effective coupling $Je^{i\phi} = J_ x + i J_y$, namely, in the $\{\ket{0}, \ket{1}\}$ basis, 
\begin{equation}
H=
\begin{pmatrix}
\Delta-i\Gamma & J\,e^{+i\phi}\\[3pt]
J\,e^{-i\phi} & 0
\end{pmatrix}. 
\label{eq:H}
\end{equation}
The non-Hermiticity, proportional to $\Gamma$, arises from the no-jump evolution of the dissipative system described in Sec.~\ref{sec:experiment}. To explore adiabatic evolution, we tune the phase of the drive $\phi(t)$ in time from 0 to $2\pi$ to form a closed loop in the parameter space. The eigenvalues $E_\pm = (\varepsilon\pm \delta)/2$ are complex, with $\varepsilon= \Delta-i\Gamma$ and $\delta/2 = \sqrt{J^2 + (\varepsilon/2)^2}=(\delta_r - i \delta_i)/2$. The right and left eigenvectors, $\ket{R_\pm}$ and $\bra{L_\pm}$, which form a bi-orthogonal basis, 
can be expressed as, 
\begin{align} 
\ket{R_{\pm}} =& \frac{1}{N_\pm} (\ket{0} + \frac{-\varepsilon\pm \delta}{2 J} e^{-i\phi} \ket{1}) \nonumber \\
\bra{L_{\pm}} =& \frac{1}{N_\pm} (\bra{0}+ \frac{-\varepsilon\pm \delta}{2 J} e^{i\phi} \bra{1}),
\label{eq:rightandleft}
\end{align}
with the normalization factors $N_\pm$ chosen such that $\braket{R_+} = \braket{R_-}=1$. Note that the square root function defining $\delta$ will be evaluated on the principal branch, without any additional phase factor. One implication is that, for $\Delta>0$, the state with the least damping is the $\ket{R_-}$ state.

The Berry connection \cite{silberstein2020berry, singhal2023measuring} follows as
\begin{equation}
A_\pm(\phi)=
i\,\frac{\bra{L_\pm}\,\partial_\phi\ket{R_\pm}}{\braket{L_\pm}{R_\pm}}
=  \frac{1}{2} \mp \frac{\varepsilon}{2\delta}.
\label{eq:berryconnection}
\end{equation}
For a trajectory on the full parameter space, see App. \ref{app:global berry phase}.

%
%
The Berry phase is obtained by integrating Eq.~\ref{eq:berryconnection} over a closed path $C_\eta$ in parameter space (Fig.~\ref{fig:fig1}b),  where $\phi$ is tuned from $0\to2\pi$, yielding
\begin{equation}
\theta_{\pm,C_{\eta}}
= \oint_{C_\eta} A_\pm(\phi)\,d\phi = 2 \pi \eta \: A_\pm
= \pi \eta\left( 1 \mp \frac{\varepsilon}{\delta}\right).
\label{eq:berryMainText}
\end{equation}
The subscripts evidence that the geometric phase depends on both the initial eigenstates and the loop direction: $\pm$ corresponds to the initial eigenstate $\ket{R_\pm}$ and $\eta=\{+,-\}$ denotes the direction of the loop ($\eta=-$ for the clockwise direction). 
 We identify that the sign of the geometric phase is (1) inverted between eigenstates, $\theta_{\pm, C_\eta} =  2\pi \eta - \theta_{\mp, C_\eta}$, and (2) flipped when the loop direction is switched: $\theta_{\pm,C_+} = -\theta_{\pm,C_-}$. 
The real part of $\theta_{\pm,C_\pm}$ includes the familiar Hermitian Berry phase, 
 while its imaginary part arises from state-dependent amplification or attenuation that are not present in normal unitary evolution.

In the Hermitian limit ($\Gamma \to 0$), the geometric phase acquired over a closed loop $C_\eta$ at resonance ($\Delta = 0$) simplifies to $\theta_{\pm, C_\eta} = \pi \eta$. This corresponds to the familiar Berry phase of $\pm \pi$ associated with a $2\pi$ rotation in parameter space, where the sign is determined by the loop direction $\eta = \pm 1$. However, in the non-Hermitian regime, the complex ratio $\varepsilon/\delta$ introduces an additional geometric component that deviates from this Hermitian baseline.


Because the closed-loop evolution must take place over a finite duration of time, the qubit will acquire a (complex) dynamical phase, $\lambda_\pm \equiv -\int_{0}^{T}E_{\pm}(t) dt$, in addition to the Berry phase. 
According to the adiabatic theorem, if the system is initially prepared in an instantaneous eigenstate $\ket{R_\pm}$ and $\phi$ is varied slowly along a closed loop $C_\eta$ over a total evolution time $T$, then the final state is: 
\begin{equation}
\ket{R_{\pm,C_\eta}} \equiv \hat{C}_\eta \ket{R_\pm} = e^{i \Theta_{\pm, C_\eta}}\ket{R_\pm},
\end{equation}
 the total phase $\Theta_{\pm, C_\eta}= \lambda_\pm + \theta_{\pm, C_\eta}$ being the sum of a dynamical phase and a geometric phase. $\hat{C}_\eta $ is the operator associated with this closed loop evolution.
For NH systems, these phases are in general complex and we specify them in terms of their real and imaginary components: 
$\lambda_{\pm} = \lambda_{\pm}^\mathrm{(r)} + i\lambda_{\pm}^{\mathrm{(im)}}$ and 
$\theta_{\pm,C_\eta} = \theta_{\pm,C_\eta}^\mathrm{(r)} + i\theta_{\pm,C_\eta}^{\mathrm{(im)}}$.
We have $\lambda_{\pm}^\mathrm{(r)}= - T(\Delta\pm \delta_r)/2$ and $\lambda_{\pm}^\mathrm{(im)}=T(\Gamma\pm \delta_i)/2$.

\section{Measurement of the Berry Phase}
\label{sec:experiment}
\subsection{Realizing non-Hermitian evolution with a dissipative superconducting qubit}


We realize an effective non-Hermitian evolution with a submanifold of a dissipative transmon circuit. The relevant energy eigenstates of the circuit are labeled $\{\ket{g}, \ket{e}, \ket{f}\}$, see Fig. \ref{fig:fig1}(a).  By engineering the circuit's dissipative environment, we establish a distinct hierarchy of decay rates:  the relaxation from the state $\ket{f}$, specified by decay rate $\gamma_f$, is suppressed relative to a comparatively rapid decay from the state $\ket{e}$, given by $\gamma_e$.   We define the differential decay parameter as the difference between these decay rates $\Gamma \equiv (\gamma_e - \gamma_f)/2 = 0.426\,\mu\mathrm{s}^{-1}$. The Hamiltonian (\ref{eq:H}) describes the dynamics within the $\{\ket{e},\ket{f}\}$ manifold when quantum jumps from $\ket{e}\to\ket{g}$ are removed via postselection. Residual dissipation within the $\{\ket{e}, \ket{f}\}$ manifold,  not captured by (\ref{eq:H}) but considered in Sec.~\ref{sec:nonunitary}, causes small perturbations on the observed dynamics \cite{chen21_jumps}.  The system is initialized in the $\{\ket{e}, \ket{f}\}$ manifold via resonant rotations on the circuit. After a period of evolution we perform a high-fidelity single-shot readout of the transmon using a SWAP operation (see \cite{lee2025nonlinear} for further details).   Via this readout, we postselect on experimental trials that remain in the $\{\ket{e}, \ket{f}\}$ manifold, thereby conditioning the evolution on the absence of quantum jumps to $\ket{g}$.

We apply a microwave drive to the circuit at frequency $\omega_d$. 
In the lab frame, the Hamiltonian can be written as $H_\textrm{lab}=-\frac{1}{2}\omega_q \sigma_z+2J \cos(\omega_dt+\phi(t))\sigma_x $. After transforming to a frame rotating at the frequency $\omega_d$ and neglecting the weak $\ket{f} \to \ket{e}$ relaxation, the conditional dynamics  are governed by the effective non-Hermitian Hamiltonian in Eq.~(\ref{eq:H}) with $\Delta\equiv\omega_d-\omega_q$.




\subsection{Interferometric measurement of the real part of the Berry phase}
\label{sec:real_part}

The real component of the geometric phase modifies the interference between instantaneous eigenstates and can be detected using an interferometric protocol \cite{leek2007observation} consisting of four steps. 

\textit{1)~Adiabatic evolution along $C_{+}$}: we initialize the system in an equal superposition of the two right eigenstates, $\ket{\psi_0}=\frac{1}{\sqrt{2}}\left(\ket{R_{+}}+\ket{R_{-}}\right)$. 
When the parameter $\phi$ is varied adiabatically along the loop $C_{+}$, the state evolves to  $\ket{\psi_1} =\hat{C}_+\ket{\psi_0}$. Since for this trajectory $\theta_{-,C_+}=2\pi-\theta_{+, C_+}$, the two components acquire opposite geometric phases, yielding
$\ket{\psi_1} =  \frac{1}{\sqrt{2}}\left(e^{i\lambda_{+}}e^{i\theta_{+, C_+}}\ket{R_{+}} + e^{i\lambda_{-}} e^{-i\theta_{+,C_+}}\ket{R_{-}}\right)$.

\textit{2)~Eigenstate swap}: to eliminate the dynamical phase contribution, we apply a resonant $\pi$ pulse along the $\mathrm{y}$ axis of the eigenbasis that exchanges $\ket{R_{+}}$ and $\ket{R_{-}}$. Namely $e^{- i \frac{\pi}{2} \hat{S}} \ket{R_\pm} = \pm \ket{R_{\mp}}$, with $\hat{S} \equiv i\ket{R_-}\bra{L_+} - i \ket{R_+}\bra{L_-}$. The resulting state is $\ket{\psi_2} =e^{- i \frac{\pi}{2} \hat{S}} \ket{\psi_1}$. 

\textit{3)~Adiabatic evolution along $C_{-}$}: We then tune $\phi$ along the reverse loop, $C_{-}$.  The final state, 
$\ket{\psi_3} =\hat{C}_- \ket{\psi_2} =  \frac{1}{\sqrt{2}}e^{i\Lambda}\left(e^{2i\theta_{+,C_+}}\ket{R_{-}}
-e^{-2i\theta_{+,C_+}}\ket{R_{+}}\right)$ 
contains an overall (complex) dynamical phase, $\Lambda = \lambda_+ + \lambda_- = - \varepsilon T$, as well as a relative phase of $4\theta_{+,C_+}$. The relative phase can be decomposed into a rotation of $4\theta^{(\mathrm{r})}_{+,C_+}$ and a relative scaling of  $e^{ -4 \theta^{(\mathrm{im})}_{+,C_+} }$. 

\textit{4)~Quantum state tomography}: we extract this rotation angle using postselected quantum state tomography. 
By defining the  observables $\hat{S}_\mathrm{x} \equiv \ket{L_-}\bra{L_+}+ \ket{L_+}\bra{L_-}$, $\hat{S}_\mathrm{y} \equiv i\ket{L_-}\bra{L_+} -i \ket{L_+}\bra{L_-}$ and the expectation values $x =  \expval{\hat{S}_\mathrm{x}}{\psi_3}$, $y = \expval{\hat{S}_\mathrm{y}}{\psi_3}$, the real geometric phase is obtained as 
$\theta_{+,C_+}^{\mathrm{(r)}} = \frac{1}{4}\arctan\!\left(y/x\right)$, providing a direct and quantitative measure of the real Berry phase accumulated during the adiabatic loop. 

More generally, for any loop direction, the above protocol the initial superposition into the state 
\begin{eqnarray}\label{eq:psietaeta}
\ket{\psi_{\eta', \eta}} \!  &=& \! \hat{C}_{\eta'} \:  e^{- i \frac{\pi}{2} \hat{S}_{y}} \: \hat{C}_{\eta}\ket{\psi_0} \\
&=& \! \frac{1}{\sqrt{2}}\Big( e^{i (\Theta_{+, \eta}{+} \Theta_{-, \eta'})} \ket{R_-}  -  e^{i (\Theta_{-, \eta}{+} \Theta_{+, \eta'})} \ket{R_+}\!\Big). \nonumber
\end{eqnarray}
When the two loop directions are the same, $\eta = \eta'$, the geometric phases cancel and $\Theta_{+, \eta}{+} \Theta_{-, \eta} = \lambda_+ + \lambda_- + 2\pi \eta$. The state simplifies to $\ket{\psi_{\eta, \eta}} = e^{i \Lambda}\ket{\psi_0}$. By contrast, opposite loop directions give
\begin{equation}
    \ket{\psi_{-\eta, \eta}} \!\! = \!\!\frac{1}{\sqrt{2}} e^{i \varepsilon T}\Big( e^{-i 2 \pi \eta \frac{\varepsilon}{\delta}}\ket{R_-}  -  e^{i 2 \pi \eta \frac{\varepsilon}{\delta}}\ket{R_+}\Big),
    \label{eq:opp_loops}
\end{equation}
with a relative phase corresponding to four times the geometric phase.

\begin{figure}
\centering
\includegraphics[width=1\columnwidth]{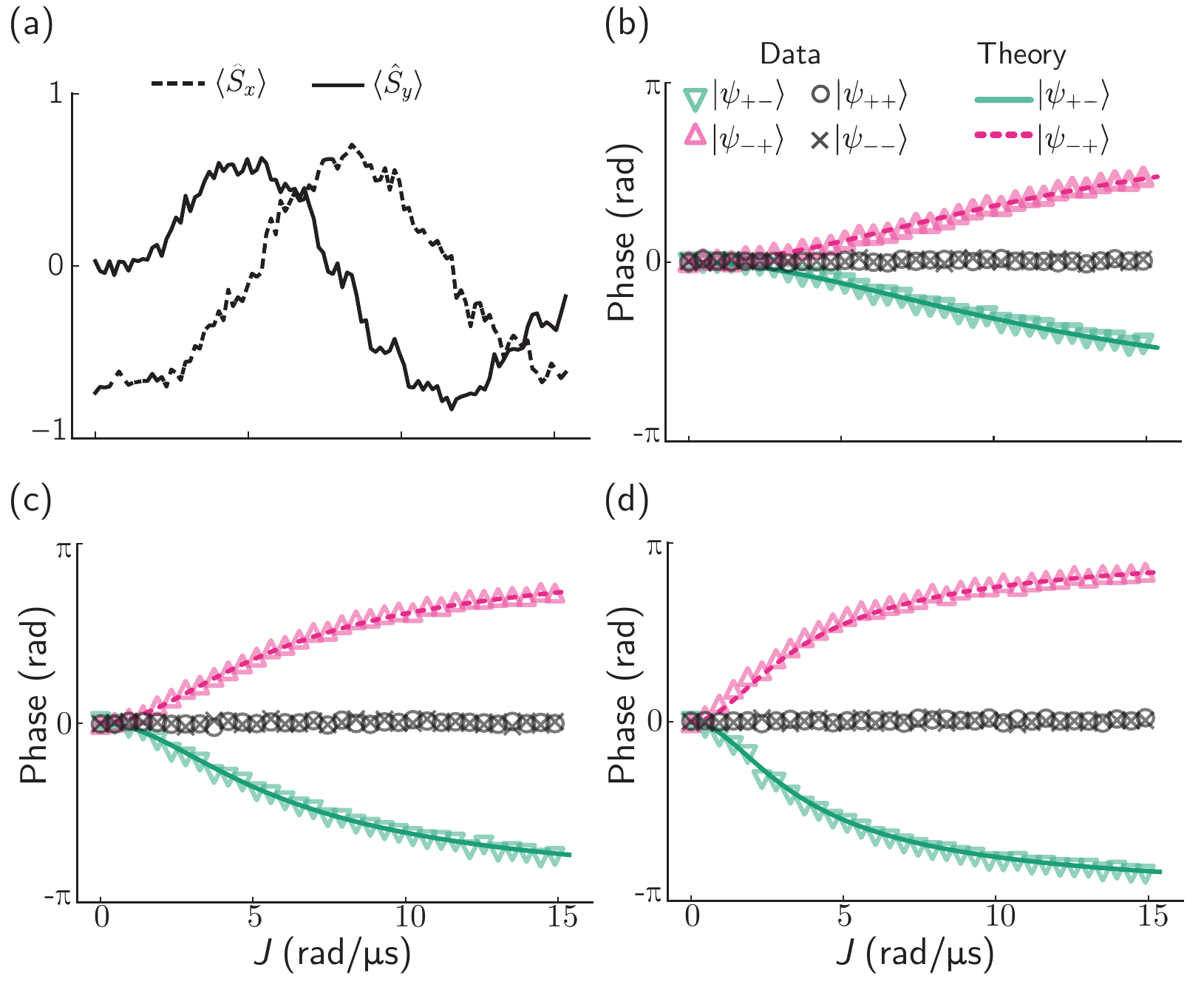}
\caption{{\bf Measurement of the real geometric phase:} 
(a) Tomographic measurement $\langle \psi_{\eta', \eta}| \hat{S}_x | \psi_{\eta', \eta}\rangle$ and $\langle \hat{S}_\textrm{y} \rangle$ at $T=3~\mu$s and $\Delta/2\pi=3$~MHz.  (b), (c) and (d) extracted real part of the geometric phase $\theta^{\textrm{(r)}}_{+,\eta}$ with $\Delta/2\pi$ = 3, 1.34, 0.81~MHz. Solid and dashed lines represent theory while markers are experimental data. Opposing loop directions ($C_{+}$ and $C_{-}$) result in sign reversal, while repeating the same loop twice with an intermediate swap ($\eta' = \eta$) cancels both geometric and dynamical phases.}
\label{fig:real_part}
\end{figure}

We apply this protocol using a single-loop duration of $T=3~\mu$s to determine the  real part of the geomeric phase, repeating the protocol for different values of $J$. Figure~\ref{fig:real_part}a shows the measured final expectation values, $x,y$ versus $J$ at fixed detuning $\Delta/2 \pi = 3$~MHz. Figure~\ref{fig:real_part}(b--d) show the extracted real part of the Berry phase for three different detunings and four different loop configurations. The phase is estimated through $\frac{1}{4}\tan(y/x)$ by fixing the principal branch at $J=0$; the branch is adjusted modulo $\pi/2$ so that the phase is analytically continued for larger couplings. The pink data points correspond to measurements performed with the sequence of loops described above, $\ket{\psi_3} = \ket{\psi_{-+}}$, 
and the green data points correspond to the opposite order of loops, $\ket{\psi_{+-}}$. 
In both cases, the experimental results show excellent agreement with the corresponding theoretical predictions (solid and dashed curves). By contrast, phase measurements performed with two identical loops, $\ket{\psi_{++}}$ and $\ket{\psi_{--}}$,
remain zero over the entire range of drive amplitudes. This is expected because the eigenstate swap causes complete cancellation of the accumulated geometric phase.

\subsection{Measurement of the imaginary part of the Berry phase}
\label{sec:imag_part}

During adiabatic evolution, the imaginary component of the Berry phase results from a change of the state norm and can therefore be extracted directly from population measurements. To probe this effect, we initialize the system in one of the instantaneous eigenstates, $\ket{R_{-}}$. 
Under an adiabatic evolution along the loop $C_{+}$, the state acquires both dynamical and geometric contributions, and becomes 
\begin{equation}
\label{eq:C_+ transform}
 \hat{C}_+ \ket{R_-} = e^{+i(\lambda_-^{\mathrm{(r)}}+\theta_{-,C_+}^{\mathrm{(r)}})}
\, e^{-(\lambda_-^{\mathrm{(im)}}+\theta_{-,C_+}^{\mathrm{(im)}})}\ket{R_{-}},
\end{equation}
where the factor $e^{-(\lambda_-^{\mathrm{(im)}}+\theta_{-,C_+}^{\mathrm{(im)}})}$ corresponds to a measurable attenuation arising from non-Hermiticity. This attenuation manifests directly as a change in the norm of the final state:
\begin{equation}
\|\hat{C}_+ \ket{R_-}\|^2 / \langle R_- | R_- \rangle
= e^{-2(\lambda_-^{\mathrm{(im)}}+\theta_{-,C_+}^{\mathrm{(im)}})} \equiv P_-^{(C_+)}.
\label{eq:rminusnorm}
\end{equation} 
In contrast to the absolute steady-state gain observed in driven-dissipative classical resonators \cite{lane2025complex, singhal2023measuring}, we measure the transient survival probability that is inherently bounded by unity. Without a continuous external drive to replenish population, the imaginary Berry phase acts as a geometric modulation of the decay rate; it provides a relative amplification that remains strictly subordinate to the dominant dynamical dissipation as $|\theta_{-, C_\eta}^{\mathrm{(im)}}|<|\lambda_-^{\mathrm{(im)}}|$, see App.~\ref{app:ImPhase}.

\begin{figure}[h]
\centering
\includegraphics[width=1.03\columnwidth]{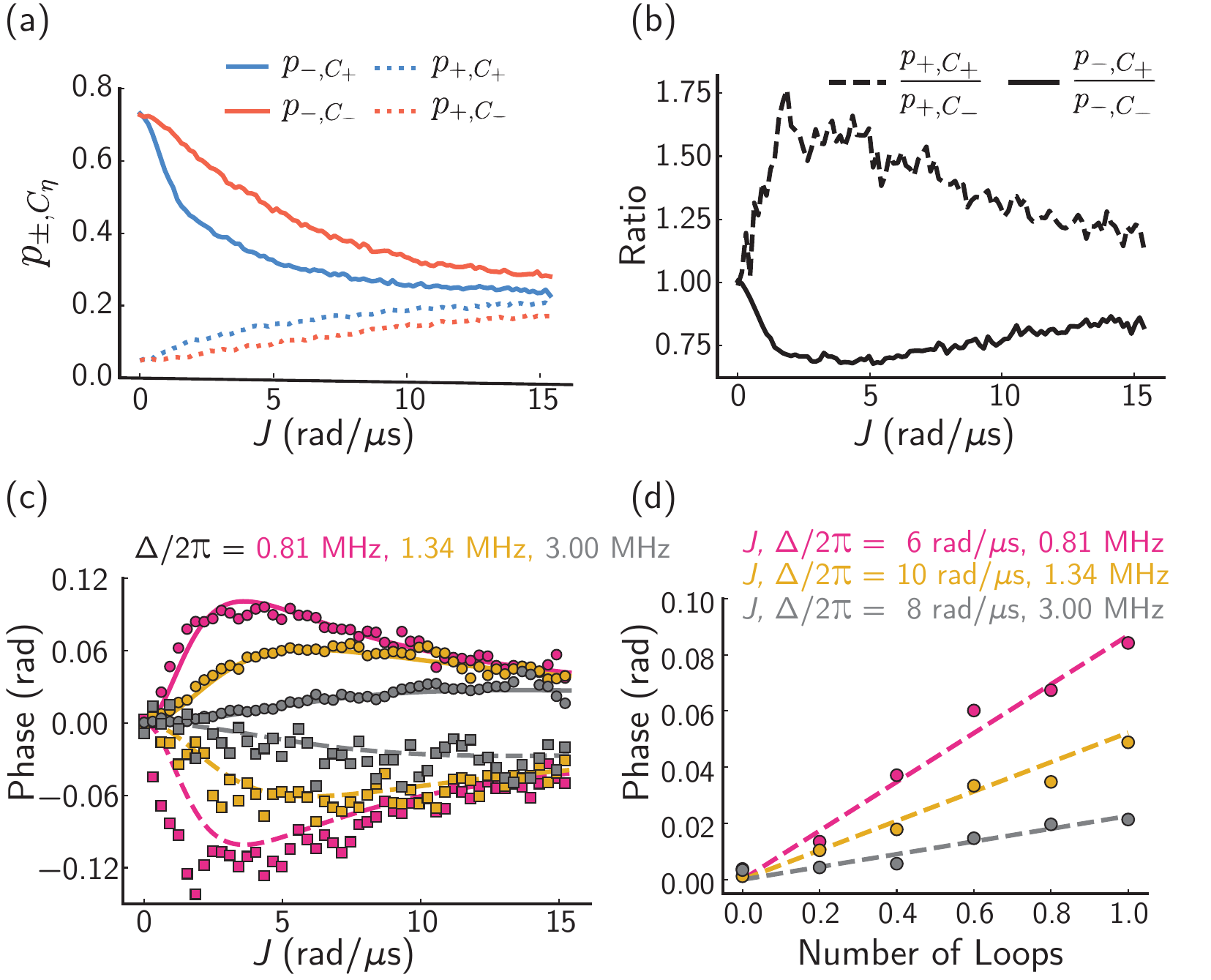}
\caption{{\bf Measurement of the imaginary part of Berry phase:} 
(a) Postselection success probability decay $p_{\pm,\eta}$ with initial state $\ket{R_+}$ (dashed) and $\ket{R_-}$ (solid) after one loop in direction $\eta$. (b) Ratio of postselection success probability obtained with eigenstates $\ket{R_+}$ (dashed) and $\ket{R_-}$ (solid) as initial states. This eliminates contribution of the dynamical phase. (c) Measured imaginary Berry phase (filled circles/squares) starting with either of the two eigenstates, compared with theoretical predictions for three different detunings. Measurements using $\ket{R_-}$ (solid lines, round markers) exhibit lower noise levels and accumulate positive phase, while those using $\ket{R_+}$ (dashed lines, square markers) show higher noise levels and accumulate negative phase. 
(d) Scaling of imaginary Berry phase in open paths for same detuning values in (c) using $\ket{R_-}$.} 
\label{fig:img_part}
\end{figure}

The reduction in the norm of the final state (in the $\{\ket{e},\ket{f}\}$ manifold) is equal to the decreasing postselection success probability (and accompanied with an increase in the population in the state $\ket{g}$). The probability $P_\pm^{(C_\eta)}$ can therefore be measured via the survival probability in the $\{\ket{e},\ket{f}\}$ manifold, denoted $p_{\pm, C_\eta}$. 
Figure~\ref{fig:img_part}a shows $p_{\pm, C_\eta}$ for a loop duration of $T=3~\mu$s for the two initial eigenstates $\ket{R_\pm}$ and loop directions ($\eta=\pm$) as function of $J$.  As is clear from Eq.~(\ref{eq:rminusnorm}), the probability depends on the imaginary part of both the geometric and dynamical phases. To eliminate the contribution from the imaginary dynamical phase, we note that repeating the same protocol along the reversed loop, $C_{-}$, leaves the dynamical evolution unchanged but reverses the sign of the geometric phase. 
Since the imaginary part of the dynamical phase contributes equally in both evolutions, it can be eliminated by taking the ratio of the two state norms, 
\begin{equation}
\frac{\| \hat{C}_+ \ket{R_-}\|^2}{\| \hat{C}_- \ket{R_-}\|^2}
 = e^{-4\theta_{-, C_+}^{\mathrm{(im)}}} = \frac{P_-^{(C_+)}}{P_-^{(C_-)}}.
 \label{eq:ratioImag}
\end{equation}
We therefore compare the ratio of $p_{\pm,C_\eta}$ for two different loop directions, as displayed in Fig.~\ref{fig:img_part}b for the two different initial states. The imaginary geometric phase can be extracted directly from this measured ratio.
Figure~\ref{fig:img_part}(c) shows this extracted imaginary part of the Berry phase versus $J$ for three different values of $\Delta$ and the two different initial eigenstates. The data show good agreement with the theoretical values (Eq.~(\ref{eq:berryMainText})) plotted as lines. 
We observe better agreement when starting in $\ket{R_-}$, the least damped eigenstate that benefits from a positive geometric Berry phase--relative amplification; whereas the evolution from $\ket{R_+}$ is only quasi-adiabatic and further attenuated, as we discuss in App.~\ref{appendix:app3}.
Furthermore, in Fig.~\ref{fig:img_part}(d), we investigate the imaginary geometric phase for the case of open loops, where $\phi:0\to 2\pi/n$, with fixed total duration $T=3~\mu$s.  We observe that $\theta_{-, C_+}^{(\mathrm{im})}$ is reduced according to the loop fraction which is a consequence of its gauge invariance: for an open loop, the eigenstate at the endpoint does not exactly coincide with the initial eigenstate (See Eq.~(\ref{eq:rightandleft})), it varies by a change in gauge. We observe that the dependence is linear.

\begin{figure}[t]
\centering
\includegraphics[width=1\columnwidth]{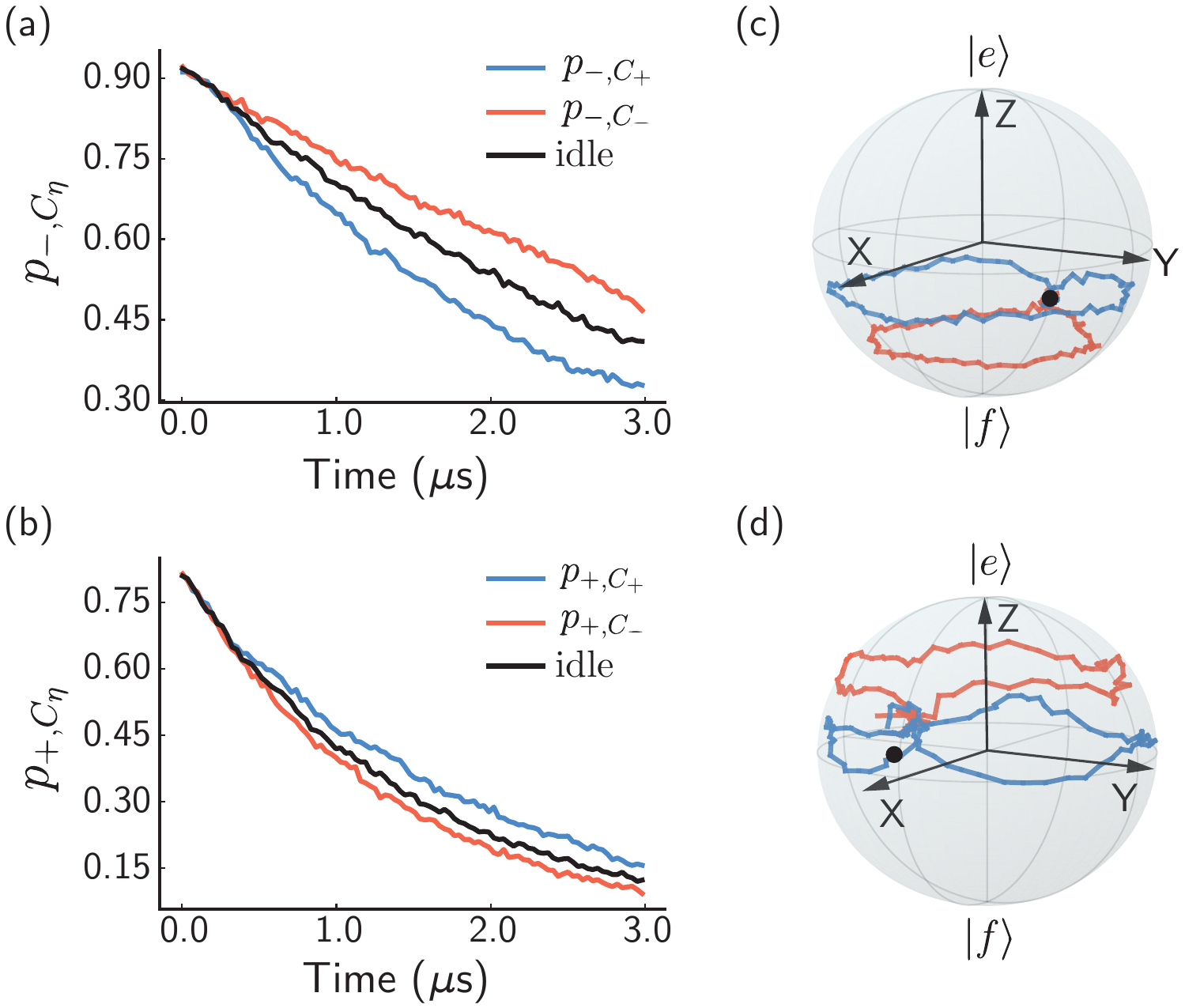}
\caption{{\bf Time dynamics of survival probability $p_{\pm, C_\eta}$ for both eigenstates under different control loop directions.} (a, b) Survival probability $p_{\pm, C_\eta}$ versus time for eigenstates $\ket{R_-}$ and $\ket{R_+}$, respectively, showing decay under control loops $C_+$ (blue), $C_-$ (red), and idle conditions (black). (c, d) Corresponding Bloch sphere trajectories for each eigenstate under loops $C_+$ and $C_-$. These trajectories are shown in the undriven frame where $\ket{R_-}\to\ket{f}$ and $\ket{R_+}\to\ket{e}$. The black dot represents the idle case initial state.}
\label{fig:fig4}
\end{figure}

Figure~\ref{fig:fig4} provides a time-resolved view of the imaginary geometric phase through the survival probability $p_{\pm, C_\eta}$ during the evolution. Relative to the idle case, which exhibits exponential decay, the control loops $C_\pm$ modify the instantaneous decay rate: for the initial state $\ket{R_-}$, the $C_-$ loop reduces the decay rate while $C_+$ enhances it. The behavior is reversed for $\ket{R_+}$. This loop-direction dependence reflects the geometric origin of the effect.

To further elucidate how the imaginary geometric phase arises, we examine the state trajectories on the Bloch sphere, displayed in Fig.~\ref{fig:fig4}(c,d). Here we display the Pauli expectation values normalized by $p$. Starting in $\ket{R_-}$, the $C_-$ ($C_+$) loop samples regions of the Bloch sphere with lower (higher) dissipation, corresponding to greater overlap with the less (more) lossy state $\ket{f}$ ($\ket{e}$), thereby reducing (enhancing) the decay rate.

The observed deviation of the instantaneous state from the ideal adiabatic trajectory provides a physical picture for the origin of the imaginary geometric phase. When the state lags towards the least damped eigenstate, it is relatively amplified; and when it lags toward the most damped eigenstate, its attenuation is increased. This lagging gives rise to the imaginary component of the geometric phase, which we formalize in App.~\ref{app:ImPhase} as being proportional to the expectation value of the anti-Hermitian part of the Hamiltonian, i.e., the difference between the actual trajectory and ideal adiabatic evolution.



\section{Non-Unitary Control using Berry Phase}
\label{sec:nonunitary}
The imaginary part of the Berry phase generates deterministic, path-dependent amplification or attenuation of the eigenstate amplitudes. This enables an additional degree of freedom for geometric operations that go beyond standard unitary control. Such non-unitary transformations could be useful in quantum simulations that require the implementation of non-Hermitian operators, including imaginary-time evolution~\cite{motta2020determining, love2020cooling}. 

\begin{figure}[b]
\centering
\includegraphics[width=\columnwidth]{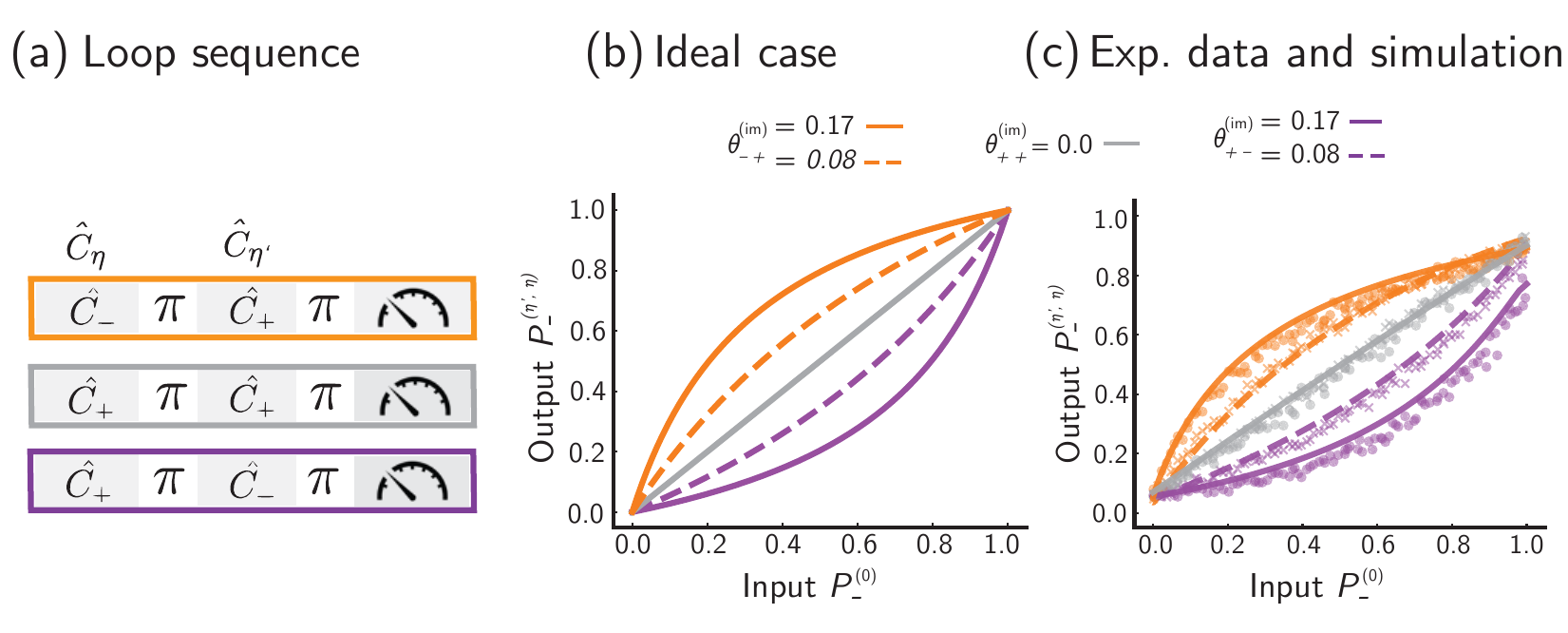}
\caption{{\bf Input–output relation for the geometric gate.} (a) Sequences with orange and purple outlines are for non-unitary control with positive and negative imaginary geometric phase, respectively. The one with gray outline corresponds to the unitary case ($\theta^{\rm (im)}_{++}=0$).  (b) Theory curves based on Eq.~(\ref{eq:echo-probs})), dashed lines are $\theta^{\mathrm{(im)}}_{+-}$ = +0.08 (orange) and $\theta^{\mathrm{(im)}}_{-+}=-0.08$ (purple) with parameters $J=4.78\,\mathrm{rad}/\mu\mathrm{s}$, $\Delta/2\pi=1.05\,\mathrm{MHz}$. Solid lines are $\theta^{\mathrm{(im)}}_{+-}$ = +0.17 (orange) and $\theta^{\mathrm{(im)}}_{-+} = -0.17$ (purple) with parameters $J=2.17\,\mathrm{rad}/\mu\mathrm{s}$, $\Delta/2\pi=0.51\,\mathrm{MHz}$. (c) Measured input-output relation for the geometric gate (colored dots and crosses). The solid and dashed curves correspond to a Lindblad simulation that captures the effect of incoherent $\ket{f} \to \ket{e}$ relaxation. Larger $|\theta^{\rm(im)}|$ produces a stronger probability–biasing effect. In experiments, the straight line corresponding to $\theta^{\mathrm{(im)}}_{\eta \eta}$ = 0 is obtained via evolution in $\hat{C_{+}}\hat{C_{+}}$ loop.}
\label{fig:inout}
\end{figure}

To demonstrate non-unitary control we apply a sequence of control loops shown in Fig.~\ref{fig:inout}. Loop operators are ordered according to the standard convention that the rightmost operator acts first. Accordingly, $\ket{\psi_{\eta',\eta}}$ denotes the state obtained after the system first undergoes the loop $\eta$, followed by the loop $\eta'$. The sequence is similar to the measurement of the real part of the Berry phase with the addition of a final $\pi$ rotation, generating the state $e^{-i\frac{\pi}{2}\hat{S}}\ket{\psi_{\eta', \eta}}$. This construction cancels the state-dependent dynamical phases, while each eigenstate accumulates opposite (complex) geometric phases. The  imaginary dynamical phase is state-independent, and can thus be
canceled out in a later normalization step.  The net action is therefore a diagonal map in the $\{\ket{R_+},\ket{R_-}\}$ basis, $U_{\mathrm{eff}} \propto \mathrm{diag}\, \!\bigl(e^{+i2\theta},\, e^{\,-i2\theta} \bigr)$, which contains both a unitary rotation ($\theta^{\mathrm{(r)}}$) and a non-unitary gain/loss factor ($\theta^{\mathrm{(im)}}$). After normalization of the final state, the transformation acts as a nonlinear map on the Bloch sphere that redistributes population toward $\ket{R_+}$ or $\ket{R_-}$ depending on the sign of $\theta^{\mathrm{(im)}}$. This constitutes a geometric non-unitary gate: the amount of gain or loss applied to each eigenbranch is set by the properties of the path in parameter space.

We analyze this explicitly by preparing a superposition in the instantaneous eigenstate basis \(\{\ket{R_+},\ket{R_-}\}\), 
\begin{equation*}
    \ket{\psi_0}=\alpha\,\ket{R_+}+\beta\,\ket{R_-},\qquad |\alpha|^2+|\beta|^2=1 .
\end{equation*}
We vary $P_{-}^{(0)}$, the initial probability of the $\ket{R_-}$ state. 
After applying the map, the final state becomes:
\begin{eqnarray}
    \ket{\psi_f} &=& e^{-i\frac{\pi}{2}\hat{S}} \ket{\psi_{\eta',\eta}} \\
    &=& e^{i \Lambda}\Big(\alpha \, e^{i2\theta_{\eta', \eta}} \ket{R_+} +\beta \, e^{-i2\theta_{\eta', \eta}}\ket{R_-}\Big),\nonumber
\end{eqnarray}
with $\theta_{\eta' ,\eta} \equiv \theta_{-,\eta'}+\theta_{+,\eta}$ equal to $2\pi \eta$ for $\eta = \eta'$ and $(- 2\pi \eta \varepsilon/\delta)$ for $\eta'= -\eta$.
The effect of the non-unitary control can be evaluated via the normalized  probabilities after the control, 
which cancel the effect of the (complex) dynamical phase and only depends on the imaginary part of the geometric phase, namely
\begin{align}
P_{-}^{(\eta', \eta)} =\frac{|\beta|^2\,e^{\,+4\theta^{\mathrm{(im)}}_{\eta', \eta}}}
       {|\beta|^2\,e^{\,+4\theta^{\mathrm{(im)}}_{\eta', \eta}}+|\alpha|^2\,e^{-4\theta^{\mathrm{(im)}}_{\eta', \eta}}},
\label{eq:echo-probs}
\end{align}
and $P_{+}^{(\eta', \eta)} = 1 - P_{-}^{(\eta', \eta)}.$
The output probabilities exhibit a nonlinear dependence on the initial superposition coefficients, $\alpha, \beta$, as captured by Eq.~\eqref{eq:echo-probs}, resulting in the relative amplitudes of $\ket{R_{+}}$ and $\ket{R_{-}}$ being rescaled by distinct exponential factors. Figure~\ref{fig:inout} summarizes an experiment that characterizes this effect. Here, we prepare different initial superpositions of $\ket{R_\pm}$ corresponding to different values of $P_{-}^{(0)}= |\beta|^2$, and then apply the control loops. We compare three types of loops show in Fig.~\ref{fig:inout}(a); a $\hat{{C}}_+ \hat{{C}}_-$ sequence that amplifies $\ket{R_-}$, a $\hat{{C}}_- \hat{{C}}_+$ sequence that attenuates $\ket{R_-}$ and finally a $\hat{{C}}_+ \hat{{C}}_+$ sequence that generates no geometric phase. Figure~\ref{fig:inout}(b) shows the transfer curves expected for an ideal non-unitary geometric gate, based on Eq.~(\ref{eq:echo-probs}). Figure~\ref{fig:inout}(c) shows the experimentally measured output probabilities. Here, the population transfer is modified by incoherent $\ket{f}\to\ket{e}$ quantum jumps as well as non-adiabaticity.  Incorporating these effects in a Lindblad simulation captures the observed behavior and yields excellent agreement with the experimental data for two different values of the imaginary geometric phase, where the strength of the non-unitary control is proportional to the magnitude of the imaginary geometric phase.

\begin{figure}[b]
  \centering
  \includegraphics[width=\columnwidth]{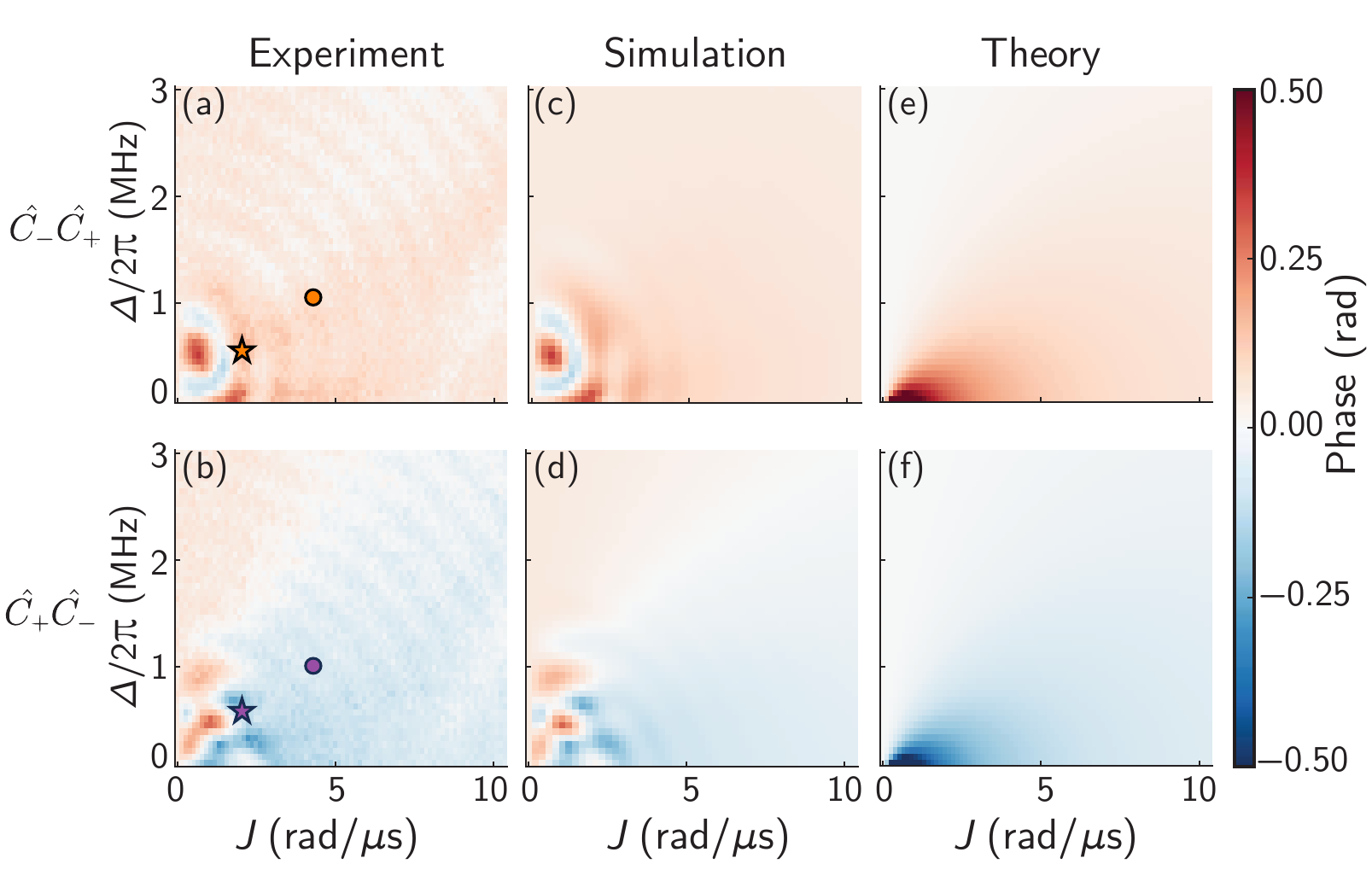}
  \caption{{\bf Imaginary part of the geometric phase for various drive amplitudes $(J)$ and detunings $(\Delta)$}: (a, b) Experiment results with $\hat{C}_- \hat{C}_+$ and $\hat{C}_+ \hat{C}_-$. The parameters $\Delta$ and $J$ used to produce orange and purple curves in Fig. 5(b) (Fig. 5(c)) are indicated with filled circles (filled stars). (c, d) Master equation simulation of same configurations including $\ket{f} \to \ket{e}$ jumps. (e, f) Theoretical predictions obtained using Eq.~\ref{eq:echo-probs}. In addition to differences in theory and experimental plots that are accounted by $\ket{f} \to \ket{e}$ jumps, the ripples in the experimental data can be attributed to diabatic evolution.
  }
  \label{fig:nonunitary_control_2D}
\end{figure}

Figure~\ref{fig:nonunitary_control_2D} summarizes how the imaginary geometric phase (and therefore strength and sign of the nonunitary control) 
can be tuned via the drive parameters $J$ and $\Delta$, the sign depending on the loop order. We display $\theta^{\mathrm{(im)}}$ as extracted from the probabilities in Eq.~\eqref{eq:echo-probs} for $\alpha = \beta = \frac{1}{\sqrt{2}}$.  
While the experimentally measured $\theta^{\mathrm{(im)}}$ [Fig.~\ref{fig:nonunitary_control_2D}(a,b)] is in good agreement with a Lindblad simulation of the full system [Fig.~\ref{fig:nonunitary_control_2D}(c,d)], the two are in marginal agreement with theory (Eq.~\ref{eq:berryMainText}), shown in [Fig.~\ref{fig:nonunitary_control_2D}(e,f)]. The deviations are attributed to $\ket{f} \to \ket{e}$ relaxation and non-adiabatic transitions, which degrade fidelity during the extended interaction times required for the adiabatic limit \aurelia{\cite{chen21_jumps}}.  
For large $J$ and large $|\Delta|$ the imaginary phase is suppressed, consistent with adiabatic following of an approximately Hermitian eigenstate. As $J$ and $\Delta$ approach the EP, the non-unitary contributions become significant \cite{comment1}. 
These maps illustrate the operational regime in which geometric control departs from purely unitary evolution, identifying the parameter regions where population biasing effects become experimentally accessible. 

\section{Outlook}\label{sec:conclude}


The Berry phase is a purely geometric contribution to quantum evolution that depends only on the path traversed in parameter space, not on the details of the trajectory. 
While complex dynamical phases arise straightforwardly from non-Hermitian eigenvalues, the complex geometric phase represents a more subtle effect: a path-dependent modulation of state amplitudes that preserves the geometric character while introducing controllable gain and loss. Our experimental demonstration establishes that both components can be independently measured and leveraged for quantum control in dissipative systems. Several promising directions emerge from this work: 

First, the imaginary geometric phase enables a new class of non-unitary gates that selectively amplify or attenuate specific eigenstates based purely on the geometry of control paths. This ``geometric filtering'' could enhance quantum state preparation protocols and enable more efficient error suppression schemes that exploit, rather than fight against, environmental dissipation. 

Second, algorithms such as quantum imaginary time~\cite{motta2020determining, love2020cooling} evolution approximate non-unitary propagators $e^{-H t}$ through sequences of variationally optimized unitary operations, a computationally intensive procedure. The eigenstate-dependent amplification demonstrated here, $|\psi\rangle = \sum_n c_n |n\rangle \to \sum_n c_n e^{-\theta^{\mathrm{(im)}}_n} |n\rangle$, naturally mimics the structure of imaginary-time evolution. Investigating whether geometric phases can reduce or eliminate the need for variational approximations represents an exciting avenue for quantum simulation.

Third, many striking phenomena in non-Hermitian physics such as topological edge states and exceptional point physics have been limited in quantum systems by the exponential cost of postselection. Our demonstration that geometric control can modulate dissipation rates suggests new strategies: by engineering parameter paths to minimize unfavorable imaginary phases, one might extend the practical regime of postselected non-Hermitian evolution and access richer phenomenology.

Finally, while our work focuses on qubit systems, the complex Berry phase should manifest equally in continuous-variable systems such as microwave resonators. 
But the interplay between geometric phases and quantum noise remains largely unexplored. Understanding how complex geometric phases govern quantum amplification and squeezing could yield new tools for quantum sensing and communication.

\section{Acknowledgments}

This research was supported by NSF Grant No.~PHY-2408932
, 
the Air Force Office of Scientific Research (AFOSR) Multidisciplinary University Research Initiative (MURI) Award on Programmable systems with non-Hermitian quantum dynamics (Grant No. FA9550-21-1-0202), 
ONR Grant No.~N000142512160, 
 and the Luxembourg National Research Fund (FNR, Attract grant QOMPET 15382998).
Devices were fabricated and provided by the Superconducting Qubits at Lincoln Laboratory (SQUILL) Foundry at MIT Lincoln Laboratory, with funding from the Laboratory for Physical Sciences (LPS) Qubit Collaboratory.


\bibliographystyle{unsrt_withcaps}
\bibliography{refs_sta}

@misc{comment1,
note = {Note that when $\Delta = 0$ and $|J|< \Gamma/2$, the imaginary part of the phase vanishes arbitrarily close to the EP. Nevertheless, in this case we still have a non-unitary contribution coming from the real part in the sense that $i\,\frac{\bra{R_\pm}\,\partial_\phi\ket{R_\pm}}{\braket{R_\pm}{R_\pm}} \neq \Re(i\,\frac{\bra{L_\pm}\,\partial_\phi\ket{R_\pm}}{\braket{L_\pm}{R_\pm}})$}
}

@article{Erdamar26,
title = {Exploring the Riemann-Surface Topology of a Non-Hermitian Superconducting Qubit Using Shortcuts to Adiabaticity},
  author = {Erdamar, Serra and Abbasi, Maryam and Chen, Weijian and H\"ornedal, Niklas and Chenu, Aur\'elia and Murch, Kater W.},
  journal = {PRX Quantum},
  volume = {7},
  issue = {1},
  pages = {010337},
  numpages = {14},
  year = {2026},
  month = {Feb},
  publisher = {American Physical Society},
  doi = {10.1103/7gtf-4tbh},
  url = {https://link.aps.org/doi/10.1103/7gtf-4tbh}
}

@article{Xu2016,
title = {Topological energy transfer in an optomechanical system with exceptional points},
volume = {537},
ISSN = {1476-4687},
url = {http://dx.doi.org/10.1038/nature18604},
DOI = {10.1038/nature18604},
number = {7618},
journal = {Nature},
publisher = {Springer Science and Business Media LLC},
author = {Xu, H. and Mason, D. and Jiang, Luyao and Harris, J. G. E.},
year = {2016},
month = jul,
pages = {80–83}
}

@article{chen21_jumps,
  title = {Quantum Jumps in the Non-{H}ermitian Dynamics of a Superconducting Qubit},
  author = {Chen, Weijian and Abbasi, Maryam and Joglekar, Yogesh N. and Murch, Kater W.},
  journal = {Phys. Rev. Lett.},
  volume = {127},
  issue = {14},
  pages = {140504},
  numpages = {6},
  year = {2021},
  month = {Sep},
  publisher = {American Physical Society},
  doi = {10.1103/PhysRevLett.127.140504},
  url = {https://link.aps.org/doi/10.1103/PhysRevLett.127.140504}
}

@article{Nagh19,
  doi = {10.1038/s41567-019-0652-z},
  url = {https://doi.org/10.1038/s41567-019-0652-z},
  year = {2019},
  month = oct,
  publisher = {Springer Science and Business Media {LLC}},
  volume = {15},
  number = {12},
  pages = {1232--1236},
  author = {M. Naghiloo and M. Abbasi and Yogesh N. Joglekar and K. W. Murch},
  title = {Quantum state tomography across the exceptional point in a single dissipative qubit},
  journal = {Nature Physics}
}

@article{Berry1984,
 ISSN = {00804630},
 URL = {http://www.jstor.org/stable/2397741},
 author = {M. V. Berry},
 journal = {Proceedings of the Royal Society of London. Series A, Mathematical and Physical Sciences},
 number = {1802},
 pages = {45--57},
 publisher = {The Royal Society},
 title = {Quantal Phase Factors Accompanying Adiabatic Changes},
 urldate = {2024-05-20},
 volume = {392},
 year = {1984}
}

@article{Dehnavi2008,
  author    = {Hossein Mehri-Dehnavi and Ali Mostafazadeh},
  title     = {Geometric Phase for Non-Hermitian Hamiltonians and Its Holonomy Interpretation},
  journal   = {Journal of Mathematical Physics},
  volume    = {49},
  number    = {8},
  pages     = {082105},
  year      = {2008},
  month     = aug,
  doi       = {10.1063/1.2967879}
}

@article{motta2020determining,
  title={Determining eigenstates and thermal states on a quantum computer using quantum imaginary time evolution},
  author={Motta, Mario and Sun, Chong and Tan, Adrian TK and O’Rourke, Matthew J and Ye, Erika and Minnich, Austin J and Brandao, Fernando GSL and Chan, Garnet Kin-Lic},
  journal={Nature Physics},
  volume={16},
  number={2},
  pages={205--210},
  year={2020},
  publisher={Nature Publishing Group UK London}
}

@article{love2020cooling,
  title={Cooling with imaginary time},
  author={Love, Peter J},
  journal={Nature Physics},
  volume={16},
  number={2},
  pages={130--131},
  year={2020},
  publisher={Nature Publishing Group UK London}
}

@article{wang2022observing,
  title={Observing a topological transition in weak-measurement-induced geometric phases},
  author={Wang, Yunzhao and Snizhko, Kyrylo and Romito, Alessandro and Gefen, Yuval and Murch, Kater},
  journal={Physical Review Research},
  volume={4},
  number={2},
  pages={023179},
  year={2022},
  publisher={APS}
}

@article{cho2019emergence,
  title={Emergence of the geometric phase from quantum measurement back-action},
  author={Cho, Young-Wook and Kim, Yosep and Choi, Yeon-Ho and Kim, Yong-Su and Han, Sang-Wook and Lee, Sang-Yun and Moon, Sung and Kim, Yoon-Ho},
  journal={Nature Physics},
  volume={15},
  number={7},
  pages={665--670},
  year={2019},
  publisher={Nature Publishing Group UK London}
}

@article{zhu2002implementation,
  title={Implementation of universal quantum gates based on nonadiabatic geometric phases},
  author={Zhu, Shi-Liang and Wang, ZD},
  journal={arXiv preprint quant-ph/0207037},
  year={2002}
}

@article{abdumalikov2013experimental,
  title={Experimental realization of non-Abelian non-adiabatic geometric gates},
  author={Abdumalikov Jr, Abdufarrukh A and Fink, Johannes M and Juliusson, Kristinn and Pechal, Marek and Berger, Simon and Wallraff, Andreas and Filipp, Stefan},
  journal={Nature},
  volume={496},
  number={7446},
  pages={482--485},
  year={2013},
  publisher={Nature Publishing Group UK London}
}

@article{arai2018geometric,
  title={Geometric phase magnetometry using a solid-state spin},
  author={Arai, Keigo and Lee, Junghyun and Belthangady, Chinmay and Glenn, David R and Zhang, Huiliang and Walsworth, Ronald L},
  journal={Nature Communications},
  volume={9},
  number={1},
  pages={4996},
  year={2018},
  publisher={Nature Publishing Group UK London}
}

@article{lin2009synthetic,
  title={Synthetic magnetic fields for ultracold neutral atoms},
  author={Lin, Y-J and Compton, Rob L and Jim{\'e}nez-Garc{\'\i}a, Karina and Porto, James V and Spielman, Ian B},
  journal={Nature},
  volume={462},
  number={7273},
  pages={628--632},
  year={2009},
  publisher={Nature Publishing Group UK London}
}

@article{zhang2005experimental,
  title={Experimental observation of the quantum Hall effect and Berry's phase in graphene},
  author={Zhang, Yuanbo and Tan, Yan-Wen and Stormer, Horst L and Kim, Philip},
  journal={nature},
  volume={438},
  number={7065},
  pages={201--204},
  year={2005},
  publisher={Nature Publishing Group UK London}
}

@article{zhou2025high,
  title={High-fidelity geometric quantum gates exceeding 99.9\% in germanium quantum dots},
  author={Zhou, Yu-Chen and Ma, Rong-Long and Kong, Zhenzhen and Li, Ao-Ran and Zhang, Chengxian and Zhang, Xin and Liu, Yang and Jiang, Hao-Tian and Wu, Zhi-Tao and Wang, Gui-Lei and others},
  journal={Nature Communications},
  volume={16},
  number={1},
  pages={7953},
  year={2025},
  publisher={Nature Publishing Group UK London}
}

@article{ozdemir2019parity,
  title={Parity--time symmetry and exceptional points in photonics},
  author={{\"O}zdemir, {\c{S}}ahin Kaya and Rotter, Stefan and Nori, Franco and Yang, L},
  journal={Nature materials},
  volume={18},
  number={8},
  pages={783--798},
  year={2019},
  publisher={Nature Publishing Group UK London}
}

@article{ding2021experimental,
  title={Experimental determination of PT-symmetric exceptional points in a single trapped ion},
  author={Ding, Liangyu and Shi, Kaiye and Zhang, Qiuxin and Shen, Danna and Zhang, Xiang and Zhang, Wei},
  journal={Physical Review Letters},
  volume={126},
  number={8},
  pages={083604},
  year={2021},
  publisher={APS}
}

@article{leek2007observation,
  title={Observation of Berry's phase in a solid-state qubit},
  author={Leek, Peter J and Fink, JM and Blais, Alexandre and Bianchetti, R and Goppl, M and Gambetta, Jay M and Schuster, David I and Frunzio, Luigi and Schoelkopf, Robert J and Wallraff, Andreas},
  journal={science},
  volume={318},
  number={5858},
  pages={1889--1892},
  year={2007},
  publisher={American Association for the Advancement of Science}
}

@article{singhal2023measuring,
  title={Measuring the adiabatic non-Hermitian Berry phase in feedback-coupled oscillators},
  author={Singhal, Yaashnaa and Martello, Enrico and Agrawal, Shraddha and Ozawa, Tomoki and Price, Hannah and Gadway, Bryce},
  journal={Physical Review Research},
  volume={5},
  number={3},
  pages={L032026},
  year={2023},
  publisher={APS}
}

@article{lane2025complex,
  title={Complex Berry phase and steady-state geometric amplification in non-Hermitian systems},
  author={Lane, JR and Guria, C and H{\"o}ller, J and Montalvo, TD and Patil, YSS and Harris, JGE},
  journal={arXiv preprint arXiv:2503.23197},
  year={2025}
}

@article{lee2025nonlinear,
  title={Nonlinear quantum evolution of a dissipative superconducting qubit},
  author={Lee, Orion and Cao, Qian and Joglekar, Yogesh N and Murch, Kater},
  journal={arXiv preprint arXiv:2510.25836},
  year={2025}
}

@article{tsubota2022symmetry,
  title={Symmetry-protected quantization of complex Berry phases in non-Hermitian many-body systems},
  author={Tsubota, Shoichi and Yang, Hong and Akagi, Yutaka and Katsura, Hosho},
  journal={Physical Review B},
  volume={105},
  number={20},
  pages={L201113},
  year={2022},
  publisher={APS}
}

@article{ruter2010observation,
  title={Observation of parity--time symmetry in optics},
  author={R{\"u}ter, Christian E and Makris, Konstantinos G and El-Ganainy, Ramy and Christodoulides, Demetrios N and Segev, Mordechai and Kip, Detlef},
  journal={Nature physics},
  volume={6},
  number={3},
  pages={192--195},
  year={2010},
  publisher={Nature Publishing Group UK London}
}

@article{bender2013observation,
  title={Observation of PT phase transition in a simple mechanical system},
  author={Bender, Carl M and Berntson, Bjorn K and Parker, David and Samuel, E},
  journal={American Journal of Physics},
  volume={81},
  number={3},
  pages={173--179},
  year={2013},
  publisher={AIP Publishing}
}

@article{schindler2012symmetric,
  title={-symmetric electronics},
  author={Schindler, Joseph and Lin, Zin and Lee, JM and Ramezani, Hamidreza and Ellis, Fred M and Kottos, Tsampikos},
  journal={Journal of Physics A: Mathematical and Theoretical},
  volume={45},
  number={44},
  pages={444029},
  year={2012},
  publisher={IOP Publishing}
}

@article{zhu2014pt,
  title={PT-symmetric acoustics},
  author={Zhu, Xuefeng and Ramezani, Hamidreza and Shi, Chengzhi and Zhu, Jie and Zhang, Xiang},
  journal={Physical Review X},
  volume={4},
  number={3},
  pages={031042},
  year={2014},
  publisher={APS}
}

@article{silberstein2020berry,
  title={Berry connection induced anomalous wave-packet dynamics in non-Hermitian systems},
  author={Silberstein, Navot and Behrends, Jan and Goldstein, Moshe and Ilan, Roni},
  journal={Physical Review B},
  volume={102},
  number={24},
  pages={245147},
  year={2020},
  publisher={APS}
}

\begin{appendix}


\section{Adiabaticity in non-Hermitian systems}
\label{appendix:app3}
In a non-Hermitian system, adiabaticity can be evaluated in terms of a transition amplitude $a_{nm}$ from state $m$ to $n$. The criterion for adiabaticity for normalized eigenvectors is given by $a_{nm}$\cite{Erdamar26}
\begin{equation}
a_{nm} = \frac{|\langle L_n(t)|\partial_t R_m(t)\rangle|}{|E_n(t) - E_m(t)|}e^{-I_{nm}(t)} \ll 1,
\label{eq_adiabatic_cond}
\end{equation}
where $I_{nm}(t)\equiv \mathrm{Im}\left[\int_{0}^{t} (E_m(t')-E_n(t'))dt'\right]$ accounts for the effect of relative gain/loss between eigenstates. In our specific case, we denote $a_{+}$ as the transition amplitude from state $|R_+\rangle$ to $|R_-\rangle$, and $a_{-}$ as the transition amplitude from $|R_-\rangle$ to $|R_+\rangle$. We calculate these amplitudes as $a_{\pm}(t) = \frac{J\dot{\phi}}{|\delta|^2}e^{\mathrm{\pm Im}(\delta)t}$.
As the total duration of the evolution approaches infinity, the phase velocity $\dot{\phi}$ approaches zero. In Hermitian systems, this is sufficient to satisfy the adiabaticity criterion. However, in the non-Hermitian case, the exponential factor dominates the behavior: it causes $a_{-}$ to vanish, while causing $a_{+}$ to diverge. This asymmetry is consistent with the breakdown of the standard adiabatic theorem observed in non-Hermitian systems.

\begin{figure}[th]
\centering
\includegraphics[width=.8\columnwidth]{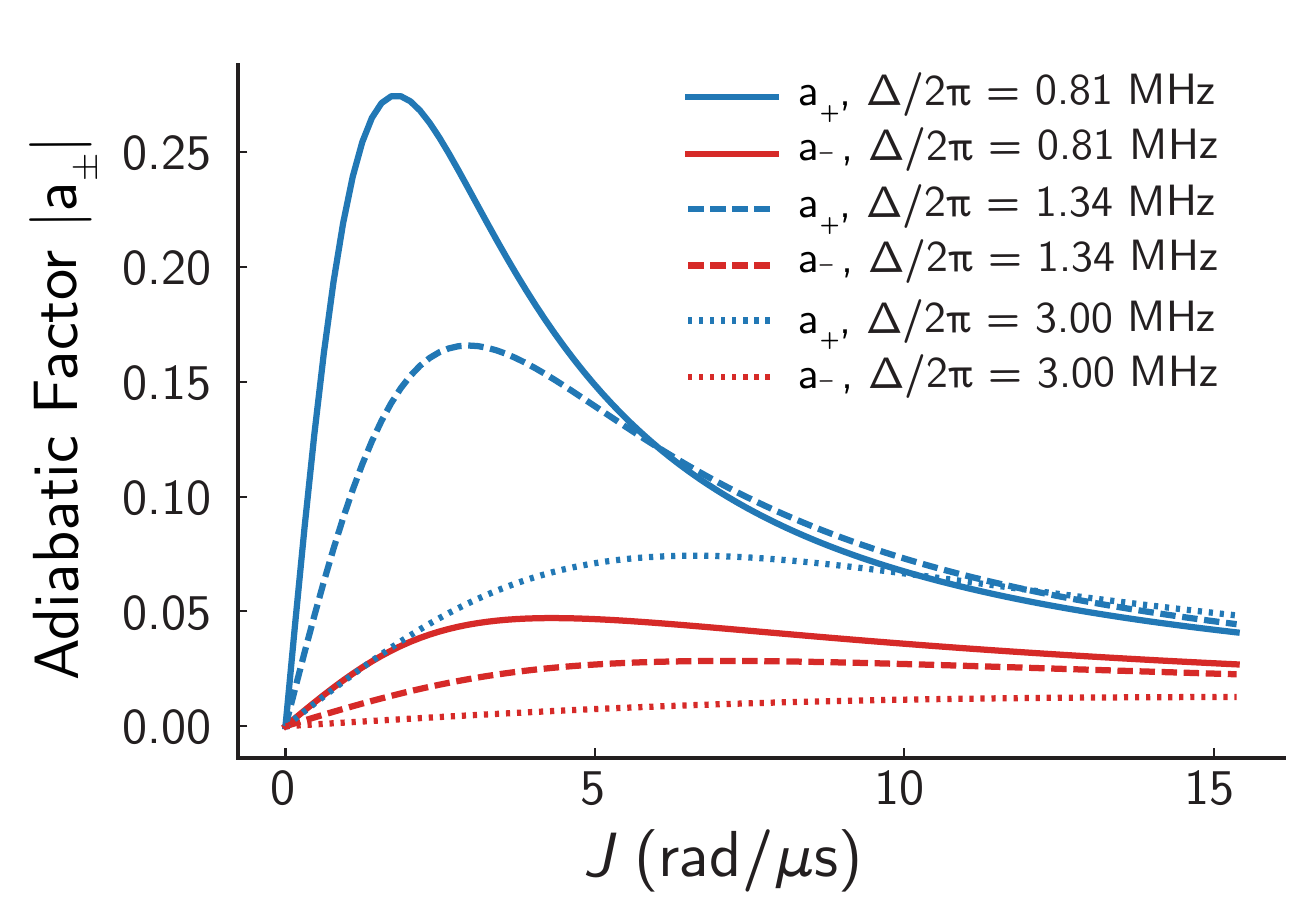}
\caption{{\bf Adibaticity of control loops.} We show the transition amplitudes $a_\pm$ for the control loops used in Sec.~\ref{sec:imag_part}. The experiments originating in $\ket{R_+}$ with transition amplitude $a_+$ are the least adiabatic.}
\label{fig:a_param}
\end{figure}

Figure~\ref{fig:a_param} displays $a_\pm$ for the control loops used in measuring the imaginary part of the geometric phase (Sec.~\ref{sec:imag_part}), where approximate $\dot{\phi}=2\pi/T$. We can see that the transition amplitude $a_+$ is in all cases larger, owing to the fact that $\ket{R_+}$ is the most damped eigenstate and can only achieve quasiadiabaiticy. However in all cases the transition amplitude $a_\pm<0.3$ corresponding to nearly adiabatic behavior. Furthermore, when applying the sweeps experimentally we find that ramping the phase velocity from zero to $\dot{\phi}$ with a cosine transition enhanced the observed adiabaiticty.  This enhances adiabaticity, especially for $a_- $ at $t=0$. For Sec.~\ref{sec:imag_part}, the cosine ramp is applied with a half-period of 750~ns and a flat phase velocity section of 1500~ns, followed by a similar cosine ramp to zero phase velocity. 
In the nonunitary control demonstration (Sec.~\ref{sec:nonunitary}) the phase velocity profile is a simple cosine pulse with a full period of 1.8 $\mu$s in Fig.~5(c) and 2.7 $\mu$s in Fig.~5(d).

\section{Geometric phase over full parameter space}
\label{app:global berry phase}

In this Appendix we provide a globally well-defined formulation of the
adiabatic geometric phase across the full parameter space.
By introducing an extended parameter space in which the eigenvectors are
single-valued, the Berry connection and curvature can be computed
explicitly. We show that the resulting non-Hermitian Berry curvature
is obtained from the Hermitian case ($\Gamma=0$) by analytic continuation,
with the mixing angle extended to complex values.

To obtain a globally well-defined description of the geometric phase,
we use the complex mixing angle defined via
\begin{equation}
\alpha
=
\frac{1}{2i}
\log\!\left(
\frac{\Delta + iJ - i\Gamma}
     {\Delta - iJ - i\Gamma}
\right),
\qquad
\alpha=\alpha_\textsc{r}+i\alpha_\textsc{i},
\end{equation}
with $\alpha_\textsc{r}\in S^1$ and $\alpha_\textsc{i}> 0$~\cite{Erdamar26}.
In terms of $(\alpha_\textsc{r},\alpha_\textsc{i})$ the physical parameters read
\begin{align}
J &= \Gamma
\frac{\cosh(2\alpha_\textsc{i}) - \cos(2\alpha_\textsc{r})}
     {\sinh(2\alpha_\textsc{i})}, \\
\Delta &= \Gamma
\frac{\sin(2\alpha_\textsc{r})}
     {\sinh(2\alpha_\textsc{i})}.
\end{align}
This parametrization defines an extended space
$\mathcal{M} = \left\{(\phi, \alpha_\textsc{r}, \alpha_\textsc{i})\in S^1\times S^1\times (0,\infty)\right\}$,
which forms a double cover of the physical $(\phi, J, \Delta)$ parameter space for $J\neq0$.
On $\mathcal{M}$ the eigenvalues and eigenvectors can be chosen
single-valued and smooth.
Consequently, any closed control loop in $\mathcal{M}$ produces a
well-defined adiabatic geometric phase without the need for gauge patching~\cite{Dehnavi2008}.

Using the above parametrization, we choose the globally smooth
biorthonormal gauge
\begin{align}
    |R_\pm\rangle &= e^{i\frac{\alpha_\textsc{r} - \phi}{2}}e^{i\frac{\phi}{2}\sigma_z}e^{-i\frac{\alpha}{2}\sigma_y}|z_\pm\rangle\\
    \langle L_\pm| &= \langle z_\pm|e^{i\frac{\alpha}{2}\sigma_y}e^{-i\frac{\phi}{2}\sigma_z}e^{i\frac{\phi - \alpha_\textsc{r}}{2}}.
\end{align}
The components of the Berry connection are given by
$A_{\pm,\mu}
=
i\langle L_\pm|\partial_{X^\mu}R_\pm\rangle$,
with parameters
$X^\mu\in\{\phi,\alpha_\textsc{r},\alpha_\textsc{i}\}$,
and one finds
\begin{align}
A_{\pm,\phi}
&=
\frac{1}{2}\bigl(1\mp\cos\alpha\bigr), \\
A_{\pm,\alpha_\textsc{r}}
&=
-\frac{1}{2}, \qquad
A_{\pm,\alpha_\textsc{i}}=0 .
\end{align}

For an experimental protocol
$t\mapsto(\phi(t), \Delta(t),J(t))$,
$t\in[0,T]$, $\alpha(t)=\alpha_\textsc{r}(t)+i\alpha_\textsc{i}(t)$ can be computed using the logarithm above. By choosing the branch continuously along the path, the geometric phase of band $\pm$ is then obtained directly as
\begin{equation}
\theta_\pm
=
\int_0^T
\left[
A_{\pm,\phi}\,\dot{\phi}
+
A_{\pm,\alpha_\textsc{r}}\,\dot{\alpha}_\textsc{r}
\right] dt ,
\end{equation}
which, using the explicit expressions, becomes
\begin{equation}
\theta_\pm
=
\int_0^T
\left[
\frac{1}{2}\bigl(1\mp\cos\alpha(t)\bigr)\dot{\phi}(t)
-
\frac{1}{2}\dot{\alpha}_\textsc{r}(t)
\right] dt .
\end{equation}
For a closed loop in $\mathcal{M}$ the second term contributes only an integer multiple of $\pi$.
This reflects the double-cover structure of the parameter space, where
the eigenvectors change sign under a $2\pi$ winding of
$\alpha_\textsc{r}$ and the geometric phase correctly captures this sign change.
The remaining contribution is given by the weighted
integral of $\dot{\phi}$ with weight
$\tfrac12(1\mp\cos\alpha)$.

Equivalently, the phase can be expressed as a surface integral 
\begin{equation}
\theta_\pm
=
\iint_D
\sum_{\mu<\nu}
F_{\pm,\mu\nu}
\left(
\partial_u X^\mu\,\partial_v X^\nu
-
\partial_v X^\mu\,\partial_u X^\nu
\right)
\,du\,dv,
\end{equation}
in terms of the gauge-independent Berry curvature components
\begin{equation}
F_{\pm,\mu\nu}
=
\partial_{X^\mu}A_{\pm,\nu}
-
\partial_{X^\nu}A_{\pm,\mu},
\end{equation}
where the only nonzero components involve $\phi$,
\begin{align}
F_{\pm,\alpha_\textsc{r}\phi}
&=
\pm\frac{1}{2}\sin\alpha, \\
F_{\pm,\alpha_\textsc{i}\phi}
&=
\pm\frac{i}{2}\sin\alpha .
\end{align}

The space $\mathcal M$ can be naturally extended by including
the boundary $\alpha_\textsc{i}=0$.
Although the parametrization of $(\Delta,J)$ degenerates there
for fixed $\Gamma\neq0$, the Berry connection and curvature remain
well-defined and smooth as functions of $\alpha$.

In the Hermitian limit ($\Gamma=0$), the parameter $\alpha$ is real,
and $(\alpha,\phi)$ coincide with the standard spherical coordinates
on the unit sphere, where $\alpha$ is the polar angle and $\phi$ the
azimuthal angle. The Berry curvature then reduces to
\[
F_{\pm,\alpha\phi}
=
\pm \frac{1}{2}\sin\alpha,
\]
which equals one half of the area element on the sphere.
Accordingly, the geometric phase equals
$\pm\tfrac12$ times the solid angle enclosed by the loop, reproducing
the familiar Bloch-sphere result for a two-level system~\cite{Berry1984}.

For $\Gamma\neq0$ the parameter $\alpha$ becomes complex
while the curvature retains the same functional dependence
$\pm\tfrac12\sin\alpha$.
Thus, the non-Hermitian curvature is obtained
by analytic continuation of the Hermitian expression
from real to complex $\alpha$.

\section{Constraint from the Imaginary Berry Phase \label{app:ImPhase}}

In this Appendix we analyze the consequences of requiring the physical state to decay at the same rate as the adiabatic solution in the presence of loss. We show that the imaginary part of the Berry connection fixes the deviation of the evolving state from the adiabatic trajectory in the dissipative ($\sigma_z$) component. In particular, the ratio $\theta_\pm^{\mathrm{(im)}}/T$ directly determines the average offset in $\langle\sigma_z\rangle$ that must be maintained for the decay rate of the physical state to match that of the adiabatic trajectory.

The adiabatic trajectory associated with the instantaneous right
eigenstate $|R_\pm(t)\rangle$ is
\begin{equation}
|\psi_{\mathrm{ad}}(t)\rangle
=
e^{-i\int_0^t E_\pm(s)\,ds}
\,e^{\,i\int_0^t A_\pm(s)\,ds}
\,|R_\pm(t)\rangle,
\end{equation}
where $E_\pm(t)$ are the instantaneous eigenvalues and
$A_\pm(t)$ is the Berry connection defined in the main text.

We recall that the anti-Hermitian part of the effective Hamiltonian is
\begin{equation}
\mathcal{D}(t)
\equiv
\frac{H_{\mathrm{eff}}^\dagger(t)-H_{\mathrm{eff}}(t)}{2i}
=
\frac{\Gamma}{2}\bigl(I+\sigma_z\bigr).
\end{equation}

Let $|\psi(t)\rangle$ denote the exact solution of the Schr\"odinger equation $\partial_t |\psi(t)\rangle
=
-i H_{\mathrm{eff}}(t)|\psi(t)\rangle$, $|\psi(0)\rangle = |\psi_{\mathrm{ad}}(0)\rangle$.
Exact adiabatic tracking,
$|\psi(t)\rangle = |\psi_{\mathrm{ad}}(t)\rangle$ for all
$t\in[0,T]$, implies equality of norms at all times,
\begin{equation}
\label{eq:scaling_condition_app}
\langle\psi(t)|\psi(t)\rangle
=
\langle\psi_{\mathrm{ad}}(t)|\psi_{\mathrm{ad}}(t)\rangle.
\end{equation}

Differentiating the norm of the evolving state gives $\partial_t \langle\psi(t)|\psi(t)\rangle
=
-2\,\langle\psi(t)|\mathcal{D}(t)|\psi(t)\rangle$. Using the adiabatic expression and imposing
Eq.~\eqref{eq:scaling_condition_app} yields
\begin{equation}
A_\pm^{\mathrm{(im)}}(t)
=
\mathrm{Tr}\!\left[
\mathcal{D}(t)\bigl(\rho(t)-\rho_{\mathrm{ad}}(t)\bigr)
\right],
\end{equation}
where $\rho(t)
= |\psi(t)\rangle\langle\psi(t)|/\langle\psi(t)|\psi(t)\rangle$ and $\rho_\mathrm{ad}(t)
= |\psi_\mathrm{ad}(t)\rangle\langle\psi(t)|/\langle\psi_\mathrm{ad}(t)|\psi_\mathrm{ad}(t)\rangle$.
Thus, the imaginary part of the Berry connection is fixed by the difference in the expectation value of anti-Hermitian part of the Hamiltonian, $\mathcal{D}$, between the evolving state and the adiabatic state.

For $\mathcal{D}=\frac{\Gamma}{2}(I+\sigma_z)$ this simplifies to
\begin{equation}
A_\pm^{(\mathrm{im})}(t)
=
\frac{\Gamma}{2}\,
\Delta z(t),
\qquad
\Delta z(t)
\equiv  z(t) - z_{\mathrm{ad}}(t),
\end{equation}
where $z(t)\equiv \mathrm{Tr}(\sigma_z\rho(t))$ and
$z_{\mathrm{ad}}(t)=\mathrm{Tr}(\sigma_z\rho_{\mathrm{ad}}(t))$.

Integrating over the protocol duration $T$ gives the imaginary part of the geometric phase
\begin{equation}
\theta_\pm^{(\mathrm{im})}
=
\frac{\Gamma}{2}
\int_0^T \Delta z(t)\,dt,
\end{equation}
and therefore
\begin{equation}
\langle\Delta z\rangle
=
\frac{1}{T}\int_0^T \Delta z(t)\,dt
=
\frac{2\,\theta_\pm^{\mathrm{(im)}}}{T\Gamma}.
\end{equation}

\begin{figure}
  \centering
  \includegraphics[width=\columnwidth]{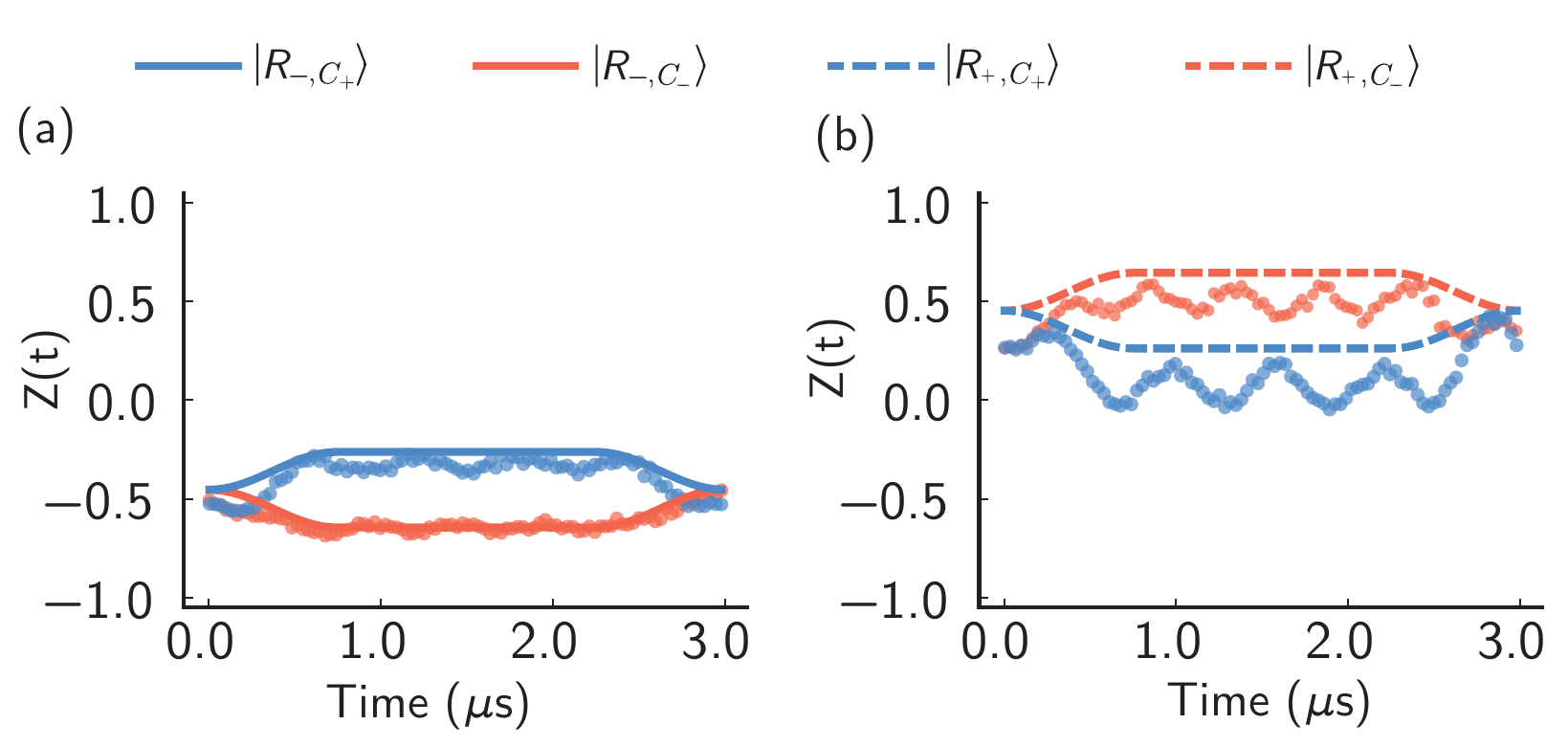}
  \caption{{\bf Comparison of experimental and theoretical $z(t)$ based on Eq.~(\ref{eq:D11})}. (a) $Z(t)$ for the $\ket{R_-}$ eigenstate. Blue and red colors indicate $C_+$ and $ C_-$ loop directions respectivley. The solid points are the experimental data from Fig.~\ref{fig:fig4}.  (b) Corresponding results for the $\ket{R_+}$ eigenstate. The oscillations in the data are due to the non-adiabatic evolution of the eigenstates and imperfect state initialization.}
  \label{fig:zt}
\end{figure}
Hence, the imaginary part of the Berry phase fixes the
time-averaged deviation in the $z$-component that the evolving state
must acquire in order to satisfy
Eq.~\eqref{eq:scaling_condition_app}. For fixed
$\theta_\pm^{\mathrm{(im)}}$, shorter protocol durations $T$ require a
larger average offset $\langle\Delta z\rangle$. In the adiabatic limit
$T\rightarrow\infty$, we recover
$\langle\Delta z\rangle \rightarrow 0$, consistent with the adiabatic
theorem.

To make this constraint explicit, we use the parametrization introduced in Appendix~\ref{app:global berry phase},
where the Berry connection component along $\phi$ is
$
A_{\pm,\phi}
=
\frac{1}{2}\bigl(1\mp\cos\alpha\bigr)$.
Its imaginary part therefore reads
\begin{equation}
\label{eq:connection_alpha}
A_\pm^{\mathrm{(im)}}(t)
=
\pm
\frac{\dot\phi(t)}{2}
\sin(\alpha_{\textsc{r}})
\sinh(\alpha_{\textsc{i}}).
\end{equation}
Integrating over the loop gives
\begin{equation}
\langle \Delta z \rangle
=
\pm
\frac{2\pi
\sin(\alpha_{\textsc{r}})
\sinh(\alpha_{\textsc{i}})}
{T\Gamma}.
\end{equation}

When $\dot z_\textrm{ad} = 0$, as is the case for a $\phi$-loop, and $A_\pm(t)$ is differentiable, we have $\dot z(t)
=
\frac{\partial_t A_\pm^{\mathrm{(im)}}(t)}{\Gamma/2}$.
Integrating from $0$ to $t$ and using $z(0)=z_{\mathrm{ad}}(0)$ then gives
\begin{equation}
z(t)
=
z_{\mathrm{ad}}(0)
+
\frac{2}{\Gamma}
\int_0^t
\partial_t A_\pm^{\mathrm{Im}}(s)\,ds.
\end{equation}
With Eq.~\eqref{eq:connection_alpha} and imposing $\dot\phi(0) = 0$, this becomes
\begin{equation}\label{eq:D11}
z(t)
=
z_{\mathrm{ad}}(0)
\pm
\frac{\sin(\alpha_{\textsc{r}})
\sinh(\alpha_{\textsc{i}})}{\Gamma}\dot\phi(t).
\end{equation}
Notably, the offset in $z(t)$ from its adiabatic value is directly proportional to the driving profile $\dot{\phi}(t)$. Fig. \ref{fig:zt} compares this prediction with the measured data presented in the main text.  

\end{appendix}
\end{document}